\documentclass[fleqn,usenatbib]{mnras}

\usepackage[T1]{fontenc}

\DeclareRobustCommand{\VAN}[3]{#2}
\let\VANthebibliography\thebibliography
\def\thebibliography{\DeclareRobustCommand{\VAN}[3]{##3}\VANthebibliography}

\usepackage{graphicx}	
\usepackage{amsmath}	
\usepackage{amssymb}	
\usepackage{gensymb}
\usepackage{caption}
\usepackage{subcaption}
\usepackage[T1]{fontenc}
\usepackage{tgbonum}
\usepackage{lscape}
\usepackage{verbatim}
\usepackage{arydshln}

\usepackage{newtxtext,newtxmath}

\newcommand\nodata{ ~$\cdots$~ }%

\title[Two new radio CVs]{Radio detections of two unusual cataclysmic variables in the VLA Sky Survey}

\author[M. Ridder et al.]{
M.~E. Ridder,$^{1}$\thanks{E-mail: mridder@ualberta.ca}
C.~O. Heinke,$^{1}$
G.~R. Sivakoff,$^{1}$
\& A.~K. Hughes$^{1}$
\\
$^{1}$ University of Alberta, 
Physics Dept., CCIS 4-183, 
Edmonton, AB 
T6G 2E1, 
Canada \\
}

\date{Accepted XXX. Received YYY; in original form ZZZ}

\pubyear{2022}

\begin{document}
\label{firstpage}
\pagerange{\pageref{firstpage}--\pageref{lastpage}}
\maketitle

\begin{abstract}
We report two new radio detections of cataclysmic variables (CVs), and place them in context with radio and X-ray detections of other CVs. We detected QS Vir, a low accretion-rate CV; V2400 Oph, a diskless intermediate polar (IP); and recovered the polar AM Her in the Very Large Array Sky Survey (VLASS) 2--4 GHz radio images. The radio luminosities of these systems are higher than typically expected from coronal emission from stars of similar spectral types, and neither system is expected to produce jets, leaving the origin of the radio emission a puzzle. The radio emission mechanism for these two CVs may be  electron-cyclotron maser emission, synchrotron radiation, or a more exotic process. We compile published radio detections of CVs, and X-ray measurements of these CVs, to illustrate their locations in the radio - X-ray luminosity plane, a diagnostic tool often used for X-ray binaries, active galactic nuclei, and radio stars. Several radio-emitting CVs, including these two newly detected CVs, seem to lie near the principal radio/X-ray track followed by black hole X-ray binaries (BHXBs) at low luminosity, suggesting additional complexity in classifying unknown systems using their radio and X-ray luminosities alone. 
\end{abstract}




\section{Introduction}

Cataclysmic variables (CVs) are binary systems in which a white dwarf (WD) accretes from an M or K type donor star \citep[e.g., see review in][]{warner_2003}. Accretion begins when the donor reaches the edge of its Roche lobe, but the accretion geometry is determined by the WD's magnetic field. If the WD is weakly magnetic, then matter from the donor fills an accretion disk around the WD. These kinds of CVs periodically go into outburst (dwarf novae) caused by an increased accretion rate, or remain at a high accretion rate for long periods of time (novalikes). The moderately magnetic case (intermediate polars; IPs) results in a truncated accretion disk, where the inner region of the disk follows the field lines onto the poles. In the most extreme case (polars), there is no accretion disk and matter only flows along the WD's magnetic field lines.

There are currently 33 known CVs detected in the radio at $\geq$3$\sigma$ confidence and 16 marginal detections. Of these 33 CVs, 23 are magnetic \citep{barrett_2020} and 10 are non-magnetic \citep{kording_2008, coppejans_2015, coppejans_2016, russell_2016}. The main hypotheses for the origin of radio emission from CVs are synchrotron emission from a jet \citep{russell_2016, mooley_2017, fender_2019}, or electron-cyclotron maser emission (ECME) near the WD or donor \citep{barrett_2020}. Even within a given class (e.g.\ novalike), there may be multiple processes responsible for the radio emission \citep{coppejans_2015,coppejans_2016}. 

SS Cyg is a well-studied non-magnetic CV with a radio counterpart \citep{russell_2016, mooley_2017, fender_2019}. Multiple radio flares coincide with its dwarf nova (DN) outbursts, and the evolution of the spectral index during one flare suggests a jet launched by the WD in SS Cyg \citep{fender_2019}. Unfortunately, there are not enough data for other DNe and novalikes to determine if their radio emission is also jet-powered \citep{coppejans_2015, coppejans_2016}. The prototypical polar, AM Her, is another well-known CV that has been observed at radio frequencies since the early 1980s \citep{Chanmugam82}. The radio source in magnetic systems such as this has been suggested to be ECME, which is the result of a plasma instability \citep{barrett_2020}. In the weakly relativistic regime, ECME produces narrow band, highly circularly polarized emission. In order to produce the instability, the plasma environment needs to be low-density \citep[$n < 10^{-12}$ \text{cm}$^{-3}$; ][]{barrett_2020} and highly magnetized, which is the case in the corona of the donor star or near the WD, possibly adjacent to the accretion column.

Among the other radio bright binaries containing white dwarfs are AR Sco \citep{marsh_2016} and AE Aqr \citep{Bookbinder87}. These are CV-like with a M-K type secondary and WD primary, but unlike CVs, there is little or no accretion, and the observed electromagnetic radiation is thought to be produced by more complex processes. AR Sco is commonly referred to as a white dwarf pulsar, as its WD's magnetic poles sweep over the  companion with a period of 1.95 minutes, which produces emission at the WD spin period and the beat period between the spin and orbital periods \citep{marsh_2016, buckley_2017, marcote_2017, garnavich_2019}. 
The leading explanation for AR Sco's emission mechanism, based on the high luminosity of the system (larger than either star should be), the beat-period signal (suggesting reprocessing of radiation from the WD striking the companion), and the rapid spindown of the WD \citep[indicating a high B field of 100--500 MG,][]{buckley_2017}, is that the emission is largely synchrotron radiation from electrons accelerated by the WD magnetic field and/or interaction between the WD and companion magnetic fields \citep[e.g.][]{marsh_2016,takata_2018}.

AE Aqr is a propeller system, meaning that as material from the donor falls towards the WD, it is flung out of the system by the magnetic field of the rapidly spinning WD \citep{Eracleous96,Wynn97}, with radio emission thought to be produced by acceleration of electrons in the outflowing material in shocks \citep{Bastian88,Meintjes03}. The only known AE Aqr twin is LAMOST J024048.51+195226.9 \citep[hereafter J0240,][]{Thorstensen2020,Garnavich2021,pretorius_2021,pelisoli_2022}. Since we do not yet know the exact process that produces radio emission in CVs, we include AE Aqr, LAMOST J0240, and AR Sco in our plots as references.

In this paper we discuss two new detections of unusual CVs in Epochs 1 and 2 of the Very Large Array Sky Survey (VLASS), QS Vir and V2400 Oph. QS Vir is a low accretion-rate ($\dot{M}=$ 1.7$\times10^{-13}$ M$_\odot$ yr$^{-1}$), eclipsing CV, sometimes referred to as a pre-CV \citep{odonoghue_2003, matranga_2012}. V2400 Oph is a diskless IP that may undergo diamagnetic blob accretion, in which blobs of matter from the donor star orbit the WD and eventually accrete or transfer back to the donor \citep{langford_2022_blob}. Based on the VLASS data, V2400 Oph has 
shown the brightest radio emission 
of any CV to date.

In Section 2 we discuss the archival radio observations of QS Vir and V2400 Oph. In Section 3 we explain the analysis of these radio images as well as archival X-ray data. In Section 4 we discuss the implications of our results. Throughout this paper, we quote errors at 1-$\sigma$ unless otherwise noted.

\section{VLASS observations}

\begin{table*}
    \centering
    \caption{Coordinates of our detected sources, from \textit{Gaia} EDR3 and in VLASS, and \textit{Gaia} distances \citep{bailer-jones_2021}. The errors quoted on VLASS distances are 1-sigma. Offsets are rounded to the nearest 0.05\arcsec.}
    \label{tab: coords}    
    \begin{tabular}{c c c c c c c}
    \hline \hline
         Name & \textit{Gaia} RA (deg) & \textit{Gaia} Dec (deg) & VLASS RA (deg) & VLASS Dec (deg) & Offset (arcsec) & Distance (pc) \\
        \hline
         QS Vir & 207.466870 & -13.226865 & 207.46694(9)& -13.22678(11) & 0.40 & 50.1$^{+0.07}_{-0.06}$ \\
         V2400 Oph & 258.151755 & -24.245759 & 258.15156(8) & -24.24585(8) & 0.75 & 700$^{+10}_{-11}$ \\
         AM Her (1.1) & 274.054911 & 49.868113 & 274.05489(11) & 49.86816(9) & 0.20 & 87.9$\pm$0.2 \\
         AM Her (2.1) & 274.054911 & 49.868113 & 274.05496(10) & 49.86810(9) & 0.10 & 87.9$\pm$0.2 \\
         \hline
    \end{tabular}
\end{table*}

The Very Large Array Sky Survey \citep[VLASS; ][]{Lacy2020} is conducted using the VLA S band (2--4 GHz) in a raster scan, with about 4.5 seconds exposure time on any source in each epoch, assuming a scan speed of 3.31''/s and a prime beam of 45/$\nu_{GHz}$. Upon completion, the full sky (above -40 declination) will be visited three times by VLASS; at the time of writing, both the first and second epochs of observations of the sky above $-40$ degrees declination are now available.
Rapid CLEAN processing by NRAO produces Quick-Look images, which have a typical noise of 120 $\mu$Jy \citep{vlass-ql}. Using the Canadian Initiative for Radio Astronomy Data Analysis\footnote{\url{https://cirada.ca/}} (CIRADA) image cutout service, an online tool that generates cutout images of radio and IR surveys, we retrieved images of each object from the list of 1681 CVs and related systems in \citet{ritter_kolb}, which were crossmatched with \textit{Gaia} EDR3 \citep{gaia_edr3}. We queried all VLASS epochs in late 2021 when observations up to Epoch 2.1 were available. We identified three radio counterparts; the polar AM Her (previously known to be radio-emitting, \citealt{Chanmugam82}), 
QS Vir, and V2400 Oph. When Epoch 2.2 was released, we checked for new observations of these three CVs, and when astrometric corrections were provided for Epoch 1, we updated our data. After astrometric corrections are applied, the absolute astrometric systematic uncertainty is $\sim$0.25\arcsec \citep{vlass_memo14}, which is better than that discussed in \citet{vlass-ql_cat}.

\begin{figure*} 
     \centering
     \begin{subfigure}[b]{0.4\textwidth}
         \centering
         \includegraphics[width=\textwidth]{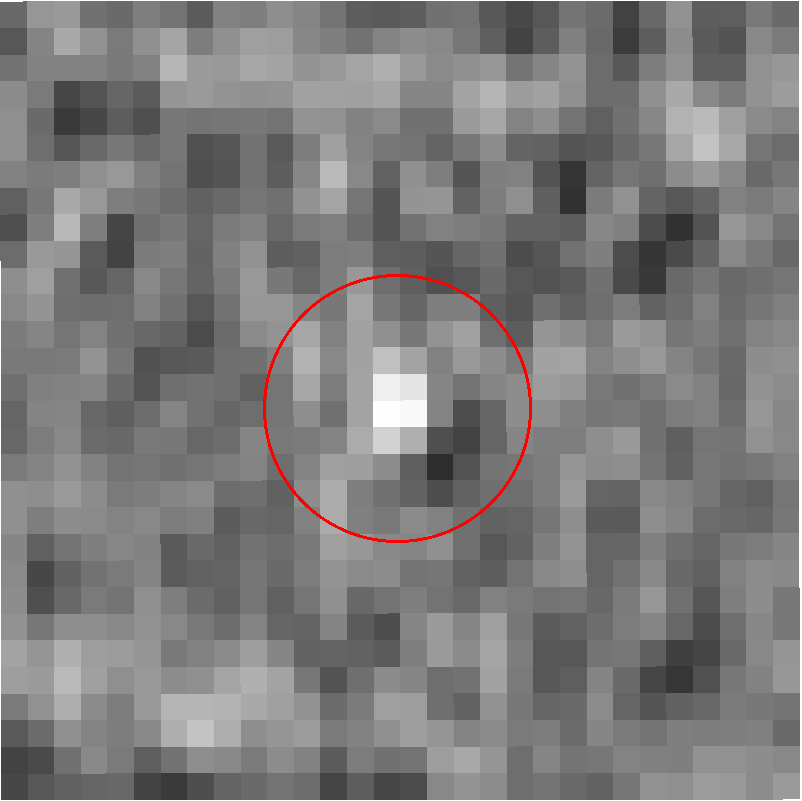}
         \caption{25 April 2019}
         \label{fig: qs vir 1.2}
     \end{subfigure}
     \begin{subfigure}[b]{0.4\textwidth}
         \centering
         \includegraphics[width=\textwidth]{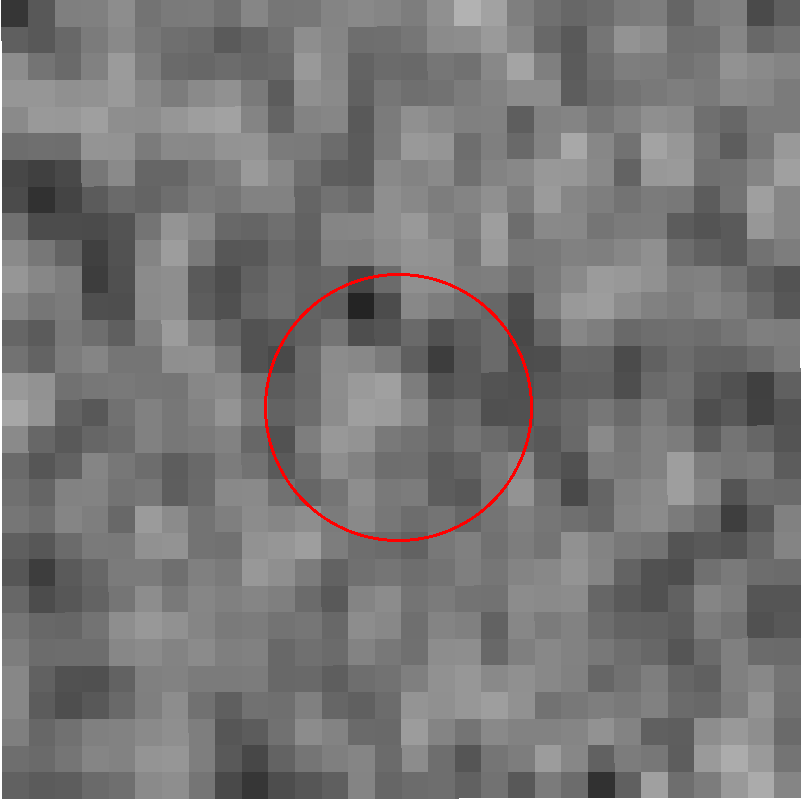}
         \caption{31 January 2022}
         \label{fig: qs vir 2.2}
     \end{subfigure}
        \caption{VLASS-QL cutouts of QS Vir on 2019 April 25 and 2022 January 31. The images are 30''$\times$30'' and the red circles have a 5 arcsecond radius, centered on QS Vir's \textit{Gaia} coordinates.}
        \label{fig: qs vir in vlass}
\end{figure*}

\begin{figure*} 
     \centering
     \begin{subfigure}[b]{0.4\textwidth}
         \centering
         \includegraphics[width=\textwidth]{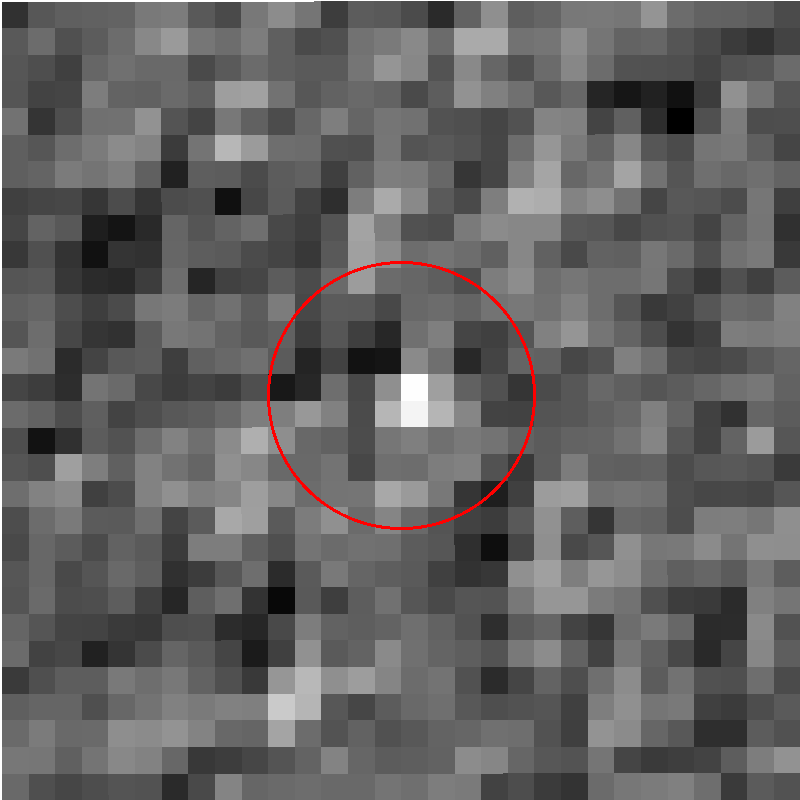}
         \caption{5 February 2018}
         \label{fig: v2400 oph 1.1}
     \end{subfigure}
     \begin{subfigure}[b]{0.4\textwidth}
         \centering
         \includegraphics[width=\textwidth]{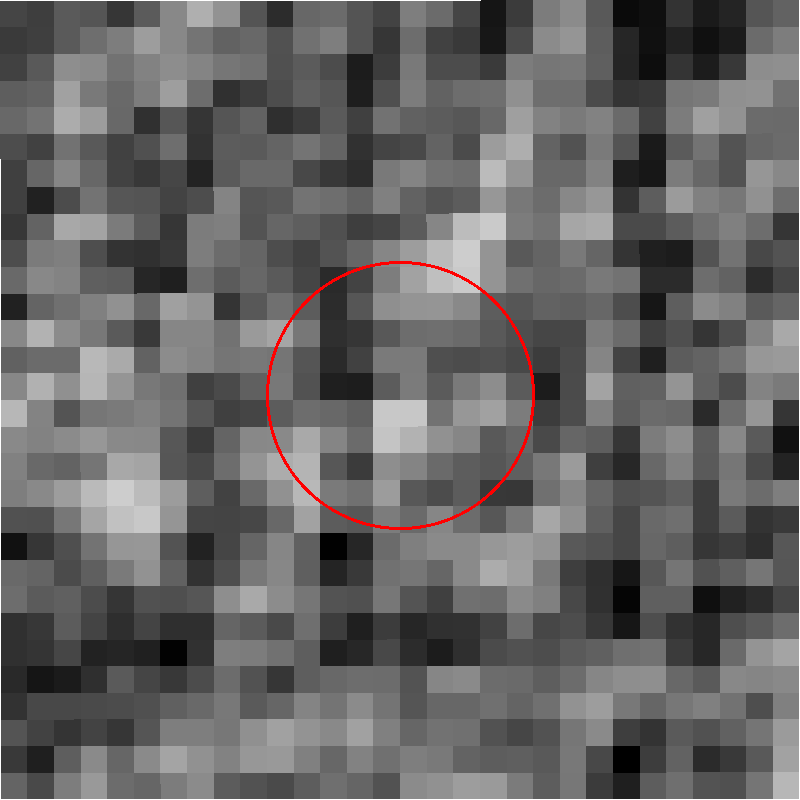}
         \caption{23 October 2020}
         \label{fig: v2400 oph 2.1}
     \end{subfigure}
        \caption{VLASS-QL cutouts of V2400 Oph on 2018 February 5 and 2020 October 23. The images are 30\arcsec$\times$30\arcsec and the red circles have a 5\arcsec 
        radius, centered on V2400 Oph's \textit{Gaia} coordinates.}
        \label{fig: v2400 oph in vlass}
\end{figure*}

\begin{figure*} 
     \centering
     \begin{subfigure}[b]{0.4\textwidth}
         \centering
         \includegraphics[width=\textwidth]{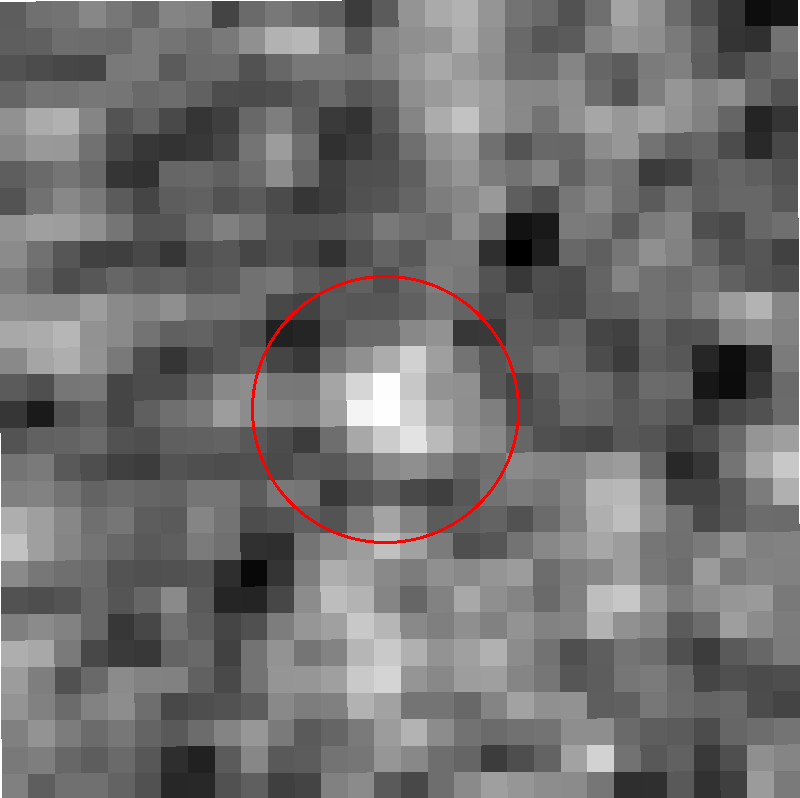}
         \caption{9 September 2017}
         \label{fig: am her 1.1}
     \end{subfigure}
     \begin{subfigure}[b]{0.4\textwidth}
         \centering
         \includegraphics[width=\textwidth]{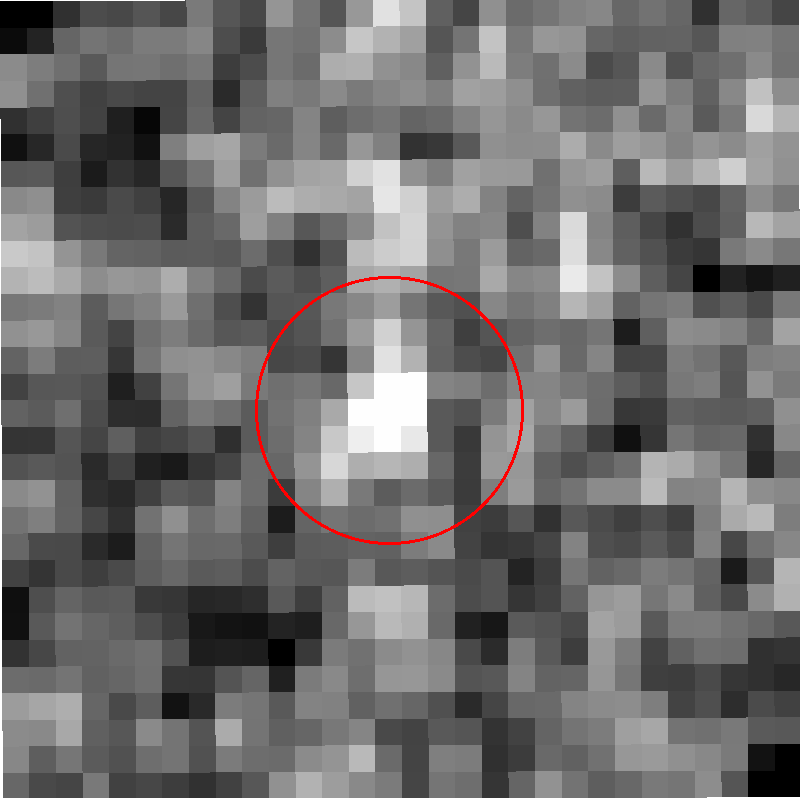}
         \caption{11 August 2020}
         \label{fig: am her 2.1}
     \end{subfigure}
        \caption{VLASS-QL cutouts of AM Her on 2017 September 9 and 2020 August 11. The images are 30''$\times$30'' and the red circles have a 5 arcsecond radius, centered on AM Her's \textit{Gaia} coordinates.}
        \label{fig: am her in vlass}
\end{figure*}

The radio images of QS Vir, V2400 Oph, and AM Her are shown in Figures \ref{fig: qs vir in vlass}, \ref{fig: v2400 oph in vlass}, and \ref{fig: am her in vlass}. Both QS Vir and V2400 Oph were detected only in the first epoch, while AM Her was detected in both epochs. As our search was performed only on the Epochs 1.1 -- 2.1, and both QS Vir and V2400 Oph appear to be transient in the radio, it seems plausible that additional CVs, including those in \cite{ritter_kolb}, may be detected in the second and third epochs of VLASS.

Based on visual inspections of the images, we do not believe the sources are likely to arise from image artifacts.
We obtained source positions and fluxes with Common Astronomy Software Applications (CASA) 6.5 \citep{casa}, software written by the National Radio Astronomy Observatory (NRAO) for analyzing radio data. The {\fontfamily{qcr}\selectfont imfit} algorithm fits a two-dimensional Gaussian distribution to a source inside a given position and radius, then returns a source position, flux, and flux error.  
To obtain the position error, we 
calculated the width of the beam 
in RA and declination. Since the synthesized beam is reported in units of the full-width-half-maximum, we divided the projected beam sizes by 2 $\times$ SNR to determine the statistical error in centroiding.
Because there is a systematic error of 1/10 of the beam in VLASS, we added this term to the last in quadrature to find the total error. We found the position of the source in Figure \ref{fig: qs vir 1.2} is 0.4\arcsec  away from the coordinates for QS Vir in \textit{Gaia} EDR3. For Figure \ref{fig: v2400 oph 1.1}, the source is 0.7\arcsec away from the \textit{Gaia} EDR3 position of V2400 Oph. The radio source near AM Her is offset by 0.2\arcsec (Figure \ref{fig: am her 1.1}) and 0.1\arcsec (Figure \ref{fig: am her 2.1}), in the two epochs respectively. 
The positions of the radio counterparts in these images are consistent within 2 $\sigma$ of the \textit{Gaia} EDR3 coordinates of each object.

With the {\fontfamily{qcr}\selectfont imfit} algorithm described above, we centered a region on each source's \textit{Gaia} position and fixed all Gaussian components to the shape of the synthesized beam for the fit. We characterized the background with the function {\fontfamily{qcr}\selectfont imstat} and estimated the noise from the root-mean-square statistic in an annulus around each source. The measured fluxes of QS Vir and V2400 Oph can be found in Table \ref{tab: flux}. 
Each CV is detected at least once at $>$5$\sigma$.

\begin{table*}
    \centering
    \caption{VLASS flux measurements (2 -- 4 GHz) of all sources when they were observed. The value for QS Vir in Epoch 2.2 and V2400 Oph in Epoch 2.1 are three sigma upper limits.}
    \label{tab: flux}    
    \begin{tabular}{c c c c c}
    \hline\hline
         Name & Flux ($\mu$Jy) & Epoch & Observation date \\
        \hline
         QS Vir & 970$\pm$160 & 1.2 & 2019 April 25 \\
         QS Vir & $\leq$ 510 & 2.2 & 2022 January 31 \\
         V2400 Oph & 640$\pm$120 & 1.1 & 2018 February 5 \\
         V2400 Oph & $\leq$ 620 & 2.1 & 2020 October 23 \\
         AM Her & 640$\pm$120 & 1.1 & 2017 September 9 \\
         AM Her & 880$\pm$150 & 2.1 & 2020 August 11 \\
         \hline
    \end{tabular}
\end{table*}

\section{Analysis}

\subsection{Chance coincidence?}
To calculate the likelihood that our radio detections are simply spurious detections of background AGN, we first need to estimate the density of background AGN. We take two approaches to this. First, we find the number of radio sources in the vicinity of QS Vir and V2400 Oph, using VLASS. Using the VLASS catalog of source from \cite{vlass-ql_cat}, there are 179 and 133 reliable components ($ Duplicate\_flag < 2$ and $Quality\_flag == 0$) from \cite{vlass-ql_cat} within a 1 degree radius around QS Vir and V2400 Oph, respectively. We thus derive an average background source density of 3.8$\times$10$^{-6}$ sources per square arcsecond.

We searched for VLASS emission near the 1681 objects from the \citet{ritter_kolb} catalog that also had \textit{Gaia} EDR3 positions.
Given typical astrometric errors in 5-$\sigma$ detected VLASS sources of 0.35\arcsec, we consider matches out to
$3\sigma$, or 1.05\arcsec. Given the 1681 potential objects, we estimate the probability of a chance coincidence at 0.022.

Since we were able to search for radio sources fainter than included in the (conservative) catalog of  \citet{vlass-ql_cat}, this chance coincidence calculation may be an underestimate. A more conservative estimate uses the expected total VLASS source number of 5.3 million anticipated detections \citep{Lacy2020}, divided by the 33885 square degrees to be surveyed. 
Since this number is for the
stacked 
combination of all three epochs, we use \citet{swire_field} to estimate that there will be about 2.3 million anticipated detections in a single VLASS Epoch, and estimate that the background source density should be about 5.2$\times$10$^{-6}$ sources per square arcsecond. Given the 1681 
error circles searched, 
we estimate the probability of a chance coincidence at 0.029.

We conclude that it is 
unlikely that either of our newly detected radio sources are chance coincidences, and that we have 
detected radio counterparts to QS Vir and V2400 Oph. 

\subsection{X-ray flux measurements}
\begin{figure}
    \centering
    \includegraphics[width=0.45\textwidth]{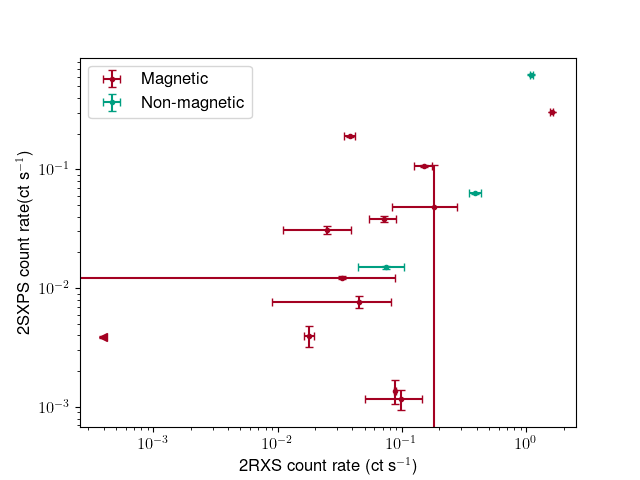}
    \caption{ROSAT (2RXS) vs. Swift (2SXPS) count rates of the objects in Tables \ref{tab: mag lr lx flux} and \ref{tab: non-mag lr lx flux}. 
    We find a typical scatter in X-ray count rates of $\sigma_X\sim 3$ ct s$^{-1}$.
    }
    \label{fig: swift/rosat}
\end{figure}

Comparison of X-ray and radio luminosities can be helpful in understanding accretion behaviour. Although we did not have (near)-simultaneous X-ray observations,
non-simultaneous X-ray observations can still be useful, especially if the X-ray flux 
is not known to  vary by orders of magnitude. 
Polars, for instance, appear to be only moderately variable, except for their  drops into "low states" with little accretion, which appear to occur about half the time \citep{Ramsay2004}. AM Her, for example, shows variability of about an order of magnitude between the Second \textit{ROSAT} All-Sky Survey Source Catalog \citep[2RXS;][]{2rxs} and the \textit{XMM-Newton} Serendipitous Source Catalog \citep[4XMM-DR11;][]{4xmm-dr11}, $\sim 10^{-13}$ and $\sim 10^{-12}$ erg s$^{-1}$ cm$^{-2}$ respectively, from 0.5 - 1 keV. However, there is more uncertainty in DNe because of the X-ray flux's dependence on whether the object is in outburst or quiescence. While the radio observations of the DNe in our dataset were taken during outburst \citep{coppejans_2016}, we cannot be certain whether the same is true for X-ray catalog data. 
Similarly, we cannot determine what accretion state the polars in our dataset were in during the X-ray or radio observations.
We estimate the X-ray variability of our sources by comparing their detections in the improved \textit{Swift XRT} Point Source Catalog \citep[2SXPS;][]{2sxps} and 2RXS (Figure \ref{fig: swift/rosat}), which allows us to estimate a dispersion of $\sim$3 ct s$^{-1}$ between multiple X-ray observations of the same source.

To obtain catalog X-ray data, we queried the HEASARC database using the \textit{Gaia} EDR3 coordinates of QS Vir and V2400 Oph, which returned detections in 4XMM-DR11. To compare to other radio-detected CVs \citep{coppejans_2015, coppejans_2016, barrett_2020}, we retrieved (generally non-simultaneous) X-ray data from their positions. These objects were observed in 4XMM-DR11, 2SXPS, the \textit{ROSAT} All-Sky Survey Bright Source Catalog \citep[RASS-BSC;][]{rassbsc}, and 2RXS. 
Only SS Cyg has published simultaneous X-ray and radio light curves during outbursts, for which we chose to plot points at the highest radio flux and the highest X-ray count rate from \citet{russell_2016}. To obtain the exact \textit{Swift} flux, we fit SS Cyg's spectra using the fitting tool provided by the UK Swift Science Data Centre\footnote{\url{https://www.swift.ac.uk/user_objects/}} \citep{swift_spectra}.

To convert count rates from these catalogs into luminosities, we searched the literature for spectral fits of magnetic CVs and non-magnetic CVs. For the few that had a spectral fit, we used the reported model and absorption column density $N_H$ to obtain the unabsorbed flux with CXC PIMMS\footnote{\url{https://cxc.harvard.edu/toolkit/pimms.jsp}}. When only catalog data were available, we used CXC Colden\footnote{\url{https://cxc.harvard.edu/toolkit/colden.jsp}} to estimate $N_H$ provided by \cite{nrao_colden_nh}. For magnetic CVs, we followed the convention in \cite{ramsay_1994}, choosing a thermal bremsstrahlung model with a 30 keV temperature for energies greater than 0.5 keV and a blackbody model with a 30 eV temperature for energies between 0.1 and 0.4 keV. In order to do this when RASS-BSC or 2RXS data were available, we used the hardness ratio between the two bands 0.1--0.4 keV and 0.5--2 keV to estimate the count rates in both. When 2SXPS data was available we used the hardness ratio between the bands 0.3--1 keV and 1--2 keV to determine whether there was a significant soft component and if so, applied a blackbody model to the 0.3--1 keV count rate. If not, a thermal bremsstrahlung model was used for that energy range. Above 1 keV, we assumed the bremsstrahlung model would dominate in all cases.

Of the novalikes in \cite{coppejans_2015} and \cite{hewitt_2020}, only TT Ari had an X-ray spectral fit \citep{mauche-mukai_2002_tt-ari}. Thus we assumed the other novalikes should be fit with a similar model and characteristic temperature; we assumed a single MEKAL model with an average temperature of 7 keV. Of the dwarf novae in \cite{coppejans_2016}, only Z Cam \citep{saitou_2012_z-cam} and SU UMa \citep{collins-wheatley_2010_su-uma} had X-ray spectral fits that could be converted with PIMMS. For these, we assumed a MEKAL model (as used in both of the fits) and an average characteristic temperature of 11 keV. 

When fluxes were available in 4XMM-DR11, we made an approximate conversion to 1--10 keV by taking the full band (Band 8, 0.2--12 keV) and subtracting two soft bands (Band 1, 0.2--0.5 keV; Band 2, 0.5--1 keV). Assuming very few photons were detected between 10 and 12 keV, the full band from 4XMM-DR11 approximates the 0.1--10 keV flux.

For data from a single band in RASS-BSC and 2RXS, we use the error given in each catalog. For data from 4XMM-DR11, we assumed the largest error quoted for the full band would be roughly the same as that for 1--10 keV. Some of the errors on \textit{Swift's} 2--10 keV and 0.3--1 keV bands were of the same magnitude so these were added in quadrature.

\subsection{Radio - X-ray luminosity plane}
We compare the radio and X-ray luminosities of the CVs, extracted from the literature as above, in Figure \ref{fig: lr-lx diagram}. The data points here are also shown in Tables \ref{tab: mag lr lx flux} and \ref{tab: non-mag lr lx flux}. We assumed a flat spectrum up to 5 GHz, 
thus multiplying the flux density by ($5\times10^9$ Hz $\times$ $4 \pi d^2$) to infer a radio luminosity. 
While some CVs have been detected at higher frequencies than 5 GHz (e.g. EQ Cet at 18-26.5 GHz), we do not have clear measurements of the shapes of CV spectra. Using a flat spectrum up to 5 GHz (while logically inconsistent for CVs detected at higher frequencies) has the benefit of allowing comparison between CVs (assuming they all have the same spectrum), and with the literature, both of CVs and X-ray binaries \citep[e.g.][]{bahramian_2018}.
We have also plotted the well-known relationship between the X-ray and radio luminosity of BHXBs  \citep[][black dotted line in Figure \ref{fig: lr-lx diagram} of this paper]{Gallo_2003}. Included in this figure are our detections, previously known radio CVs, and the unique systems, AR Sco and AE Aqr. V2400 Oph stands out in the upper right corner, far brighter in radio and X-ray than most magnetic and non-magnetic CVs. The second brightest CV in the radio is V1323 Her, another IP \citep{barrett_2020}. QS Vir is near the bottom left, near several magnetic CVs. For comparison of our data to X-ray binaries, we have plotted a transitional millisecond pulsar \citep[tMSP; PSR J1023+0038;][]{bahramian_2018, deller_2015_psr, bogdanov_2018_psr} and three BHXBs \citep[A0620-00, XTE J1118+480, Cen X-4; ][]{bahramian_2018, fender_2010_bhxb, gallo_2014_bhxb, tudor_2017_cenx4}.

Sources in \cite{barrett_2020} 
with radio detections below 3-$\sigma$ (Cas 1, FL Cet, Hya 1, HS0922+1333, and Her 1) were excluded in this plot. 
We plotted the highest and lowest radio fluxes, for objects with multiple radio detections, as a range denoted with a dashed line.

\subsection{Radio luminosity and orbital period}
\begin{figure*}
    \centering
    \includegraphics[width=0.75\textwidth]{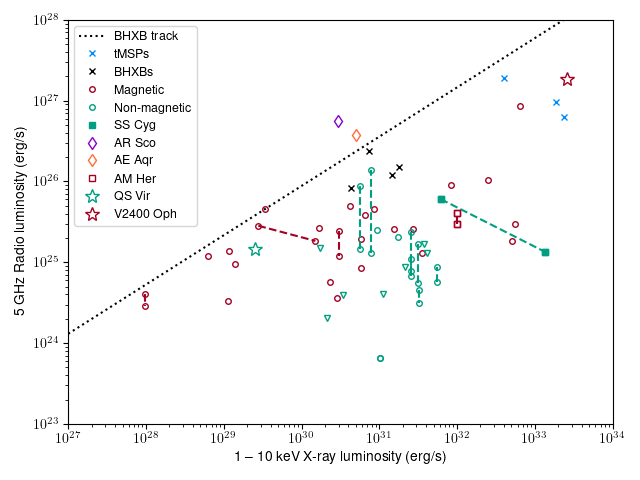}
    \caption{X-ray vs.\ radio luminosities of CVs (and related systems) with detections in both bands. Two locations of simultaneous data (filled symbols) are plotted for SS Cyg, one at maximum X-ray luminosity and one at maximum radio luminosity \citep{russell_2016}. All other data are non-simultaneous (open symbols). Multiple radio detections of the same object at different luminosities are connected with dashed lines. The black dotted line marks the L$_X$-L$_R$ relationship for BHXBs \citep{Gallo_2003}.}
    \label{fig: lr-lx diagram}
\end{figure*}

\begin{figure}
    \centering
    \includegraphics[width=0.45\textwidth]{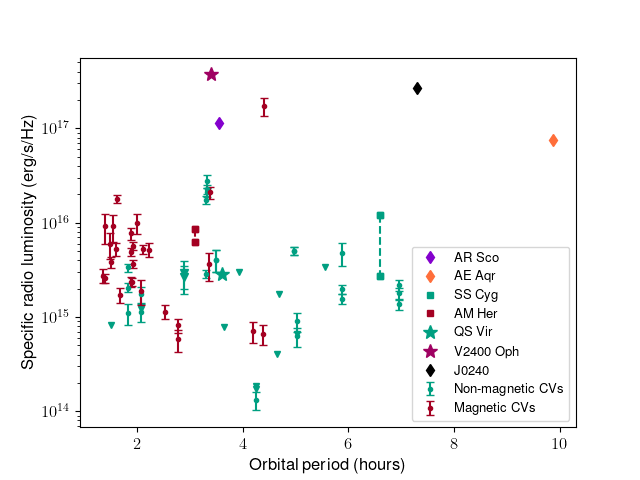}
    \caption{The orbital period in hours for our sample vs.\ specific radio luminosity. Here we plot each individual observation of magnetic and non-magnetic CVs in \citet{barrett_2020,coppejans_2015,coppejans_2016,hewitt_2020}, unlike Figure \ref{fig: lr-lx diagram}. LAMOST J0240 is an AE Aqr twin that we do not show in Figure \ref{fig: lr-lx diagram} since it currently is undetected in X-rays.}
    \label{fig: period vs lr}
\end{figure}

Figure \ref{fig: period vs lr} shows our detections compared to previously known radio CVs, AR Sco, AE Aqr, and the candidate propeller system LAMOST J0240 (note that the latter is not yet X-ray-detected, and thus not shown in Figure \ref{fig: lr-lx diagram}). V2400 Oph's orbital period is much shorter than the propellers, but it is similarly bright, as is the other IP, V1323 Her. QS Vir is within the cluster of magnetic and non-magnetic CVs, including AM Her.

\begin{figure}
    \centering
    \includegraphics[width=0.45\textwidth]{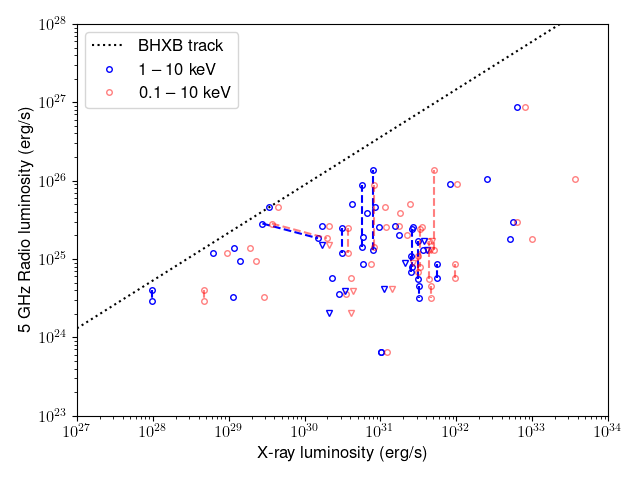}
    \caption{Comparing the effects of using the 1--10 keV X-ray luminosity vs. the 0.1--10 keV X-ray luminosity for the positions of radio-detected CVs in the radio/X-ray plane.
    The black dotted line marks the L$_X$-L$_R$ relationship for BHXBs \citep{Gallo_2003}.}
    \label{fig: soft v hard}
\end{figure}

\section{Discussion}
\begin{figure}
    \centering
    \includegraphics[width=0.45\textwidth]{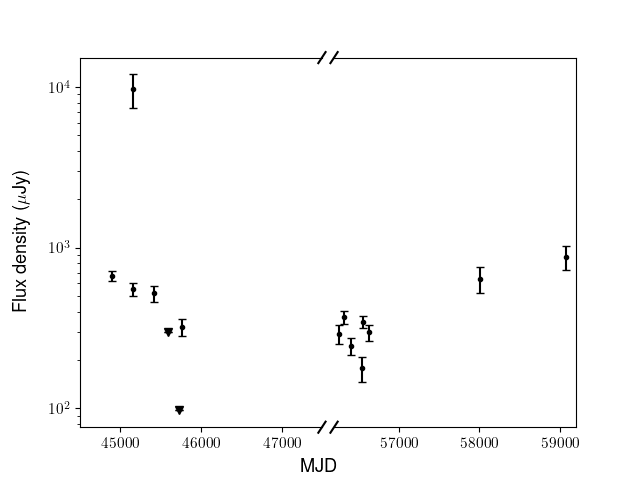}
    \caption{4.9 GHz radio observations of AM Her from 1981 to 2013 (MJD 44900 -- 56600) \citep{amher_flare_1983, amher_radio_1985, amher_radio_2018} and the 2 -- 4 GHz VLASS observations from 2017 and 2020 (MJD 58000 -- 59100). Upper limits are at the 1-$\sigma$ level. For the majority of its history, the radio flux varies by a factor of $\sim$ 4; the outlying points are the 9.7 $\pm$ 2.3 mJy flare and lower limit of 100 $\mu$Jy.}
    \label{fig: amher radio}
\end{figure}

The proximity of our detections and the rest of the CV population to the hard-state BH X-ray binary relationship in Figure \ref{fig: lr-lx diagram} suggests a possible problem for radio/X-ray identification of quiescent BHXBs. The radio-X-ray relation has often been used to suggest that a given radio source near the BHXB line is likely to be a BHXB \citep[e.g.][]{Bahramian2020,Zhao2020,Shishkovsky2018}. However, our results show that there could be significant CV contamination among sources near this line. We note that CVs often show substantial X-ray variability of an order of magnitude or more, and, based on the limited observations so far, radio variability seems common at well. Simultaneous X-ray and radio observations would be very useful. We provide a series of radio observations of AM Her in Figure \ref{fig: amher radio} to illustrate the observed variability on a timescale of years. Except for a flare of 9.7 $\pm$ 2.3 mJy \citep{amher_flare_1983} and a 100 $\mu$Jy upper limit detected by \cite{amher_radio_1985}, the flux varies between $\sim$200 to $\sim$700 $\mu$Jy. However, as most of our data is non-simultaneous, interpreting Figure \ref{fig: lr-lx diagram} should be done with caution.

Given the limited VLASS data at present, we cannot be certain of the emission mechanism behind the radio counterparts of V2400 Oph and QS Vir. In the case of QS Vir, the extremely low accretion rate paired with the known lack of an  accretion disc suggests radio emission originates in ECME rather than jets. Radio flares on isolated M dwarfs are fairly common, and \cite{hallinan_2008} propose that the dominant radio emission mechanism in late type dwarfs is ECME. 
QS Vir has displayed coronal flares from its M-type donor star in the past \citep{odonoghue_2003}, suggesting the radio flares could be from the donor. However, 
\citet{gudel_2002} shows that M and K dwarf radio detections in the cm band generally reach $10^{15}$ erg/s/Hz at most, which for 5 GHz gives a maximum  $L_R=5\times10^{23}$ erg/s; a factor of 20 below QS Vir's detection. 
This does not prove that the radio emission cannot come from the donor star, as there is a candidate detection of brighter radio emission from a young M-type brown dwarf (2MASS J16044075–1936525) in \citet{Ling2022}, at $L_R$(5 GHz)=$1.1\times10^{27}$ erg/s. However, such radio emission is quite rare; \citet{Ling2022} surveyed 2600 (mostly young) stellar objects and detected only 6 of them in radio. 
\citet{Ling2022} produced a stacked 5 GHz radio luminosity upper limit of $<1.3\times10^{23}$ erg/s for the RECONS sample (821 M stars within $\sim$70 pc from Earth,  \citealt{Henry2018}).
Similarly, \citet{Launhardt2022} observed 170 young stars at 6 GHz, finding a half dozen detections near $10^8$ W/Hz, or $L_R$(5 GHz)=$5\times10^{24}$ erg/s, but only one subgiant (V875 Per) brighter than QS Vir.

The source of the radio emission in V2400 Oph is even more difficult to explain. Its larger (by a factor of 100) radio luminosity (larger than any known M or K dwarf) would make ECME from the companion very surprising, while the lack of an accretion disc in the system makes the jet hypothesis unlikely. \citet{langford_2022_blob} suggest that the accretion flow in V2400 Oph consists of a series of disconnected diamagnetic blobs \citep{King1993, Littlefield2021}, which may either ride down the magnetic field lines to the magnetic poles of the white dwarf (generating the clearly observed spin periodicity), or gain energy from the magnetic field and be expelled, likely back to the donor star. This suggests the possibility of a similarity in the radio emission mechanism between V2400 Oph and the propeller systems AE Aquarii and J0240, which also expel material and have similar radio luminosity to V2400 Oph. On the other hand, V2400 Oph has not shown evidence for $\sim$3000 km/s motion of propeller-launched blobs, as seen in AE Aquarii and J0240. Alternatively, \citet{Kennedy2020} identify evidence of high-velocity (600--1000 km/s) features during a low state of the IP FO Aquarii, which they suggest may be attributed to a jet or fast outflow.  IPs lack a disk near the WD surface, which makes the formation of a jet in such systems challenging to justify theoretically. If this is indeed ECME, the source could be akin to the model for polars in \cite{barrett_2020}, near the short-lived accretion stream caused by the blobs.

Future work on these CVs will include longer radio observations of these systems to measure the spectral index, polarization, and (with simultaneous X-ray observations) place the system firmly on the Lx/Lr plot. 
We do not know if there is a pattern to the radio  flares, nor how frequently they recur, so 
a series of several 1-hour 
observations on different days
may be the best option to catch their radio emission and advance our understanding of these unusual systems.

\begin{table*}
    \setlength{\tabcolsep}{1.5pt}
    \centering
    \caption{The fluxes for magnetic CVs used to generate Figure \ref{fig: lr-lx diagram} and \ref{fig: soft v hard}, including AR Sco, AE Aqr, AM Her, and V2400 Oph. Refer to Table \ref{tab: non-mag lr lx flux} for non-magnetic CVs.}
    \label{tab: mag lr lx flux}
    \begin{tabular}{l c c c c c c c c c}
    \hline\hline
        Name           & RA                            & Dec                   & Distance                 & 1--10 keV flux               & 0.1--10 keV flux               & Radio band          & Radio flux & X-ray ref. & Radio ref. \\
                       & (deg)                         & (deg)                 & (pc)                     & (erg s$^{-1}$ cm$^{-2}$)    & (erg s$^{-1}$ cm$^{-2}$)      & (GHz)               &  ($\mu$Jy) &            &    \\
        \hline
        EQ Cet         & \phantom{0}22.21913           & $-23.66237$           & 283 $^{+5}_{-6}$         & 9$\pm4\times10^{-13}$       & 1.2$\pm0.4\times10^{-12}$     & 18 – 26.5        & 96$\pm30$       & 9  & 2 \\
        BS Tri         & \phantom{0}32.37423           & \phantom{$-$}28.54133 & 277 $^{+8}_{-7}$         & 1.7$\pm0.6\times10^{-12}$   & 1.9$\pm0.6\times10^{-12}$     & 8 – 12           & 57$\pm9$        & 22 & 2 \\
        EF Eri         & \phantom{0}48.5543\phantom{0} & $-22.5948$\phantom{0} & 163 $^{+66}_{-50}$       & 3.7$\times10^{-14}$         & 6.0$\times10^{-14}$           & 8 – 12           & 87$\pm15$       & 19 & 2 \\
        UZ For         & \phantom{0}53.86947           & $-25.73939$           & 238 $\pm 3$              & 2.5$\pm0.7\times10^{-13}$   & 3.1$\pm0.7\times10^{-13}$     & 4 – 8            & 78$\pm9$        & 20 & 2 \\
        RXJ0502.8+1624 & \phantom{0}75.71247           & \phantom{$-$}16.40595 & 217 $^{+13}_{-10}$       & 1.0$\pm0.2\times10^{-10}$   & 1.1$\pm0.2\times10^{-10}$     & 8 – 12           & 105$\pm32$      & 3  & 2 \\
        LW Cam         & 106.04185                     & \phantom{$-$}62.05777 & 549 $\pm 19$             & 2.3$\pm0.5\times10^{-12}$   & 2.8$\pm0.5\times10^{-12}$     & 8 – 12           & 50$\pm5$*       & 22 & 2 \\ 
        VV Pup         & 123.77841                     & $-19.05524$           & 135.3 $\pm 0.7$          & 2.7$\times10^{-12}$         & 3.5$\times10^{-12}$           & 8 – 12           & 79$\pm14$       & 4  & 2 \\
        FR Lyn         & 133.55839                     & \phantom{$-$}39.09354 & 480 $^{+50}_{-40}$       & 2.4$\pm1.1\times10^{-13}$   & 6.5$\pm1.1\times10^{-13}$     & 8 – 12           & 28$\pm4$        & 22 & 2 \\
        WX LMi         & 156.61466                     & \phantom{$-$}38.75055 & 96.8 $^{+0.5}_{-0.6}$    & 8.7$\times10^{-15}$         & 4.2$\times10^{-14}$           & 4 – 8            & 73$\pm12$       & 23 & 2 \\
        \nodata        & \nodata                       &\nodata                & \nodata                  & \nodata                     & \nodata                       & 8 - 12           & 52$\pm14$       &\nodata&\nodata\\
        ST LMi         & 166.41570                     & \phantom{$-$}25.10783 & 114.2 $\pm 0.9$          & 4.0$\pm0.5\times10^{-14}$   & 6.0$\pm0.6\times10^{-14}$*    & 8 – 12           & 153$\pm15$*     & 24 & 2 \\ 
        AR UMa         & 168.93529                     & \phantom{$-$}42.97289 & 98.6 $\pm 0.5$           & $2.4\times10^{-13}$        & $3.2\times10^{-13}$          & 4 – 8            & 489$\pm49$*     & 21 & 2 \\ 
        \nodata        & \nodata                       & \nodata               & \nodata                  & $1.3\times10^{-12}$        & $1.7\times10^{-12}$          & 18 – 26.5        & 317$\pm32$*     &\nodata&\nodata\\ 
        EU UMa         & 177.48200                     & \phantom{$-$}28.75199 & 287 $^{+16}_{-15}$       & 6$^{+2}_{-5}\times10^{-13}$ & \nodata                       & 8 – 12           & 39$\pm5$        & 14 & 2 \\
        V1043 Cen      & 198.32121                     & $-32.98694$           & 172.9 $\pm 0.9$          & 7.9$\pm0.8\times10^{-13}$*  & 1.0$\pm0.1\times10^{-12}$*       & 8 – 12           & 20$\pm5$        & 24 & 2 \\ 
        J1503-2207     & 225.97494                     & $-22.11956$           & 387$\pm$ 18              & 1.5$\pm0.2\times10^{-12}$   & 2.0$\pm0.2\times10^{-12}$     & 8 -- 12          & 29$\pm$5        & 9  & 2 \\
        BM CrB         & 235.26945                     & \phantom{$-$}36.04798 & 423 $^{+14}_{-16}$       & 1.6$\pm1.5\times10^{-14}$   & 2.1$\pm1.5\times10^{-14}$     & 8 – 12           & 43$\pm15$       & 24 & 2 \\
        MR Ser         & 238.19645                     & \phantom{$-$}18.94162 & 131.2 $^{+0.5}_{-0.7}$   & $1.5\times10^{-12}$         & $1.8\times10^{-12}$           & 4 – 8            & 239$\pm24$*     & 13 & 2 \\ 
        \nodata        & \nodata                       & \nodata               & \nodata                  & \nodata                     & \nodata                       & 8 – 12           & 116 $\pm 15$    &\nodata&\nodata\\
        MQ Dra         & 238.37941                     & \phantom{$-$}55.27067 & 180.7 $^{+2.1}_{-1.7}$   & 2.9$\pm0.8\times10^{-14}$   & 7.5$\pm0.8\times10^{-14}$     & 8 – 12           & 17 $\pm 4$      & 24 & 2 \\
        AP CrB         & 238.55136                     & \phantom{$-$}27.36463 & 199.6 $^{+1.8}_{-2.1}$   & 4.9$\pm1.6\times10^{-13}$   & 8.7$\pm1.6\times10^{-13}$     & 8 – 12           & 24 $\pm 4$      & 22 & 2 \\
        \textbf{AR Sco} & 245.44706                    & $-22.88645$           & 116.5 $^{+0.4}_{-0.5}$   & $1.80\times10^{-12}$        & \nodata                       & 5.5              & 7000$^{+5000}_{-2000}$ & 11&11 \\
        \textbf{V2400 Oph} & 258.15176                 & $-24.24576$           & 700$^{+10}_{-11}$        & 4.4$\pm0.4\times10^{-11}$*  & \nodata                       & 2 – 4            & 640$\pm$120     & 24 & 25 \\ 
        V1007 Her      & 261.02625                     & \phantom{$-$}41.23535 & 470 $^{+40}_{-30}$       & 1.6$\pm0.7\times10^{-13}$   & 9.4$\pm0.9\times10^{-13}$*    & 8 – 12           & 38 $\pm 9$      & 22 & 2 \\ 
        V1323 Her      & 270.91524                     & \phantom{$-$}40.20558 & 1830 $^{+180}_{-140}$    & 1.6$\pm0.2\times10^{-12}$*  & 2.0$\pm0.2\times10^{-12}$*    & 4 – 8            & 43 $\pm 9$      & 24 & 2 \\
        \textbf{AM Her}& 274.05491                     & \phantom{$-$}49.86811 & 87.9 $\pm 0.2$           & 1.1$\pm0.1\times10^{-10}$*  & \nodata                       & 2 – 4            & 640 $\pm 120$   & 24 & 25 \\
        \nodata        & \nodata                       & \nodata               & \nodata                  & \nodata                     & \nodata                       & \nodata          & 880$\pm 150$    &\nodata&\nodata\\
        V1432 Aql      & 295.04755                     & $-10.42382$           & 450 $\pm 7.0$            & $2.1\times10^{-11}$         & $4.1\times10^{-11}$           & 8 – 12           & 15 $\pm 5$      & 15 & 2 \\
        J1955+0045     & 298.80199                     & \phantom{$-$}0.76011 & 165.5 $^{1.9}_{-1.5}$    & 1.1$\pm0.1\times10^{-11}$   & 1.3$\pm0.1\times10^{-11}$     & 8 -- 12          & 79$\pm$ 8*      & 24 & 2 \\
        \textbf{AE Aqr}& 310.03848                     & $-0.87079$            & 91.34 $^{+0.12}_{-0.13}$ & 5.1$\pm0.5\times10^{-12}$*  & \nodata                       & 4.9              & 7500$\pm750$*   & 8  & 1 \\ 
        HU Aqr         & 316.99219                     & $-5.29488$            & 189.6 $\pm$ 1.5          & $3.3\times10^{-14}$         & $5.4\times10^{-14}$           & 8 – 12           & 44 $\pm 13$     & 18 & 2 \\
        V388 Peg       & 329.38469                     & \phantom{$-$}8.92079  & 720 $^{+80}_{-50}$       & 4.1$\pm0.8\times10^{-12}$   & 6$\pm2\times10^{-11}$   & 8 – 12           & 34 $\pm 5$      & 9  & 2 \\
        \hline
        \multicolumn{10}{p{0.9\textwidth}}{Note. --- The coordinates and distances in this table were found in Gaia EDR3 and \citet{bailer-jones_2021} respectively, except for EF Eri, which did not appear in that catalog. That information was taken from \citet{thorstensen_2003_ef-eri-dist}. RXJ0502.8+1624 is equivalent to Tau 4 in \citet{barrett_2020}. Where there are multiple entries for the same object, the row will begin with a ``\nodata'' in the ``Name'' column and anywhere the values are the same as the previous row. Otherwise, a ``\nodata'' indicates the information is unavailable, such as for our detections and the highlighted objects in Figure \ref{fig: lr-lx diagram}, since those were not included in Figure \ref{fig: soft v hard}. In the case of EU UMa, the model from \citet{ramsay_2004_eu-uma} did not provide enough information to find the flux between 0.1 $-$ 10 keV, so that entry was left blank. Errors on the X-ray and radio flux columns are assumed to be at least 10\% and were raised (and marked with an asterisk) if quoted as less in their references. The references in the final two columns are numbered as follows: 
    [1] \citet{abada-simon_1993_aeaqr_radio}, 
    [2] \citet{barrett_2020}, 
    [3] \citet{2rxs}, 
    [4] \citet{bonnet-bidaud_2020_vv-pup}, 
    [5] \citet{collins-wheatley_2010_su-uma}, 
    [6] \citet{coppejans_2015}, 
    [7] \citet{coppejans_2016},  
    [8] \citet{eracleous_1991_aeaqr_x-ray}, 
    [9] \citet{2sxps}, 
    [10] \citet{hewitt_2020}, 
    [11] \citet{marsh_2016}, 
    [12] \citet{mauche-mukai_2002_tt-ari}, 
    [13] \citet{ramsay_1994}, 
    [14] \citet{ramsay_2004_eu-uma}, 
    [15] \citet{rana_2005_v1432-aql}, 
    [16] \citet{russell_2016} 
    [17] \citet{saitou_2012_z-cam}, 
    [18] \citet{schwarz_2009_hu-aqr}, 
    [19] \citet{schwope_2007_ef-eri}, 
    [20] \citet{still_2001_uz-for}, 
    [21] \citet{szkody_1999_ar-uma}, 
    [22] \citet{rassbsc}, 
    [23] \citet{vogel_2007_wx-lmi}, 
    [24] \citet{4xmm-dr11}, 
    [25] this paper.}
    \end{tabular}
\end{table*}

\begin{table*}
    \setlength{\tabcolsep}{1.5pt}
    \centering
    \caption{The fluxes for non-magnetic CVs used to generate Figure \ref{fig: lr-lx diagram} and \ref{fig: soft v hard}, including QS Vir and SS Cyg. Refer Table \ref{tab: mag lr lx flux} for magnetic CVs and the list of citations in the table note.}
    \label{tab: non-mag lr lx flux}    
    \begin{tabular}{c c c c c c c c c c}
    \hline\hline
        Name           & RA                  & Dec                   & Distance               & 1--10 keV flux                     & 0.1--10 keV flux               & Radio band          & Radio flux & X-ray ref. & Radio ref. \\
                       & (deg)               & (deg)                 & (pc)                   & (erg s$^{-1}$ cm$^{-2}$)          & (erg s$^{-1}$ cm$^{-2}$)      & (GHz)               &  ($\mu$Jy) &            &    \\
        \hline
        CM Phe         & \phantom{00}5.38842 & $-51.70967$           & 308$\pm3$              & 3.3$\pm$0.3$\times10^{-12}$*      & 4.2$\pm$0.4$\times10^{-12}$   & 0.43 – 2.1       & < 30            & 9 & 10 \\ 
        RX And         & \phantom{0}16.14809 & \phantom{$-$}41.29928 & 196.6 $\pm 0.9$        & 7$\pm2\times10^{-13}$             & 1.0$\pm0.1\times10^{-11}$*    & 8 – 12           & 19.6 $\pm 4.4$  & 9  & 7 \\ 
        \nodata        & \nodata             & \nodata               &\nodata                 & \nodata                           & \nodata                       & \nodata          & 13.6 $\pm 3.2$  & \nodata&\nodata\\
        TT Ari         & \phantom{0}31.72112 & \phantom{$-$}15.29490 & 247 $\pm 2$            & $2.40\times10^{-11}$              & $3.10\times10^{-11}$          & 4 – 8            & 39.6 $\pm 4.2$  & 12 & 6 \\
        \nodata        & \nodata             & \nodata               & \nodata                & \nodata                           & \nodata                       & \nodata          & 240 $\pm 24$*   &\nodata&\nodata\\ 
        IM Eri         & \phantom{0}66.17152 & $-20.12003$           & 185.5 $^{+1.0}_{-0.9}$ & 4.3$\pm1.1\times10^{-12}$         & 5.5$\pm1.0\times10^{-12}$     & 0.43 – 2.1       & 99 $\pm 26$     & 3  & 10 \\
        V347 Pup       & \phantom{0}92.64028 & $-48.74036$           & 290.3$\pm$1.2          & 1.7$\pm$0.4$\times10^{-13}$       & 2.1$\pm$0.6$\times10^{-13}$   & 0.43 – 2.1       & < 30            & 9 & 10 \\
        U Gem          & 118.77167           & \phantom{$-$}22.00122 & 93.1 $\pm 0.3$         & 1.0$\pm0.1\times10^{-11}$*        & 1.2$\pm0.1\times10^{-11}$*    & 8 – 12           & 12.7 $\pm 2.8$  & 24 & 7 \\ 
        YZ Cnc         & 122.73616           & \phantom{$-$}28.14233 & 233.5 $^{+1.5}_{-1.8}$ & 8.5$\pm0.8\times10^{-12}$*        & 1.5$\pm0.2\times10^{-11}$*    & 8 – 12           & 17.4 $\pm 3.7$  & 24 & 7 \\ 
        \nodata        & \nodata             & \nodata               & \nodata                & \nodata                           & \nodata                       & \nodata          & 26.8 $\pm 5.2$  &\nodata&\nodata\\
        IX Vel         & 123.82884           & $-49.22283$           & 90.13$\pm$0.14         & 2.2$\pm$0.2$\times10^{-12}$*      & 4.2$\pm$0.4$\times10^{-12}$*  & 0.43 – 2.1       & < 42            & 24 & 10 \\ 
        SU Uma         & 123.11785           & \phantom{$-$}62.60612 & 220.1 $\pm 1.1$        & $5.40\times10^{-12}$              & $7.60\times10^{-12}$          & 8 – 12           & 19.1 $\pm 4.9$  & 5  & 7 \\ 
        \nodata        & \nodata             & \nodata               & \nodata                & \nodata                           & \nodata                       & \nodata          & 58.1 $\pm 5.8$* & \nodata  & \nodata \\ 
        Z Cam          & 126.30486           & \phantom{$-$}73.11083 & 213.5 $^{+1.2}_{-1.3}$ & 4.7$\pm0.5\times10^{-12}$*        & 5.7$\pm0.6\times10^{-12}$*    & 8 – 12           & 25.0 $\pm 3.1$  & 17 & 7 \\ 
        \nodata        & \nodata             & \nodata               & \nodata                & \nodata                           & \nodata                       & \nodata          & 40.3 $\pm 5.2$  &\nodata&\nodata\\
        RW Sex         & 154.98584           & $-8.69899$            & 221.6 $^{+1.1}_{-1.0}$ & 4.4$\pm0.5\times10^{-12}$         & 5.6$\pm0.6\times10^{-12}$*    & 0.86 – 1.7       & 233 $\pm 36$    & 9  & 10 \\ 
        \nodata        & \nodata             &\nodata                & \nodata                & \nodata                           & \nodata                       & 4 – 8            & 26.8 $\pm 3.3$  & \nodata & 2 \\
        \textbf{QS Vir} & 207.46687          & $-13.22687$           & 50 $^{+0.07}_{-0.06}$  & 8.5$\pm0.9\times10^{-13}$*        & \nodata                       & 2 – 4            & 970$\pm$160     & 24 & 25 \\ 
        V1084 Her      & 250.94041           & \phantom{$-$}34.04427 & 465 $\pm 4$            & 1.6$\pm0.2\times10^{-12}$*        & 1.8$\pm0.2\times10^{-12}$*    & 4 – 8            & < 10.2 $\pm3.4$ & 24 & 6 \\ 
        LS IV -08 3    & 254.12335           & \phantom{0}$-8.57738$ & 211$\pm2$              & 4.0$\pm$0.7$\times10^{-12}$       & 5.1$\pm$0.9$\times10^{-12}$   & 0.43 – 2.1       & < 33            & 3 & 10 \\
        V341 Ara       & 254.42249           & $-63.21105$           & 155.3$\pm$0.8          & 1.2$\pm$0.2$\times10^{-12}$       & 1.5$\pm$0.3$\times10^{-12}$   & 0.43 – 2.1       & < 27            & 9  & 10 \\
        V603 Aql       & 282.22770           & \phantom{$-$}0.58408  & 315 $^{+3}_{-4}$       & 6.7$\pm1.1\times10^{-13}$         & 4.3$\pm0.4\times10^{-12}$*    & 4 – 8            & 22 $\pm$ 3      & 9  & 6 \\ 
        \nodata        & \nodata             & \nodata               & \nodata                & \nodata                           & \nodata                       & 0.86 – 1.7       & 233 $\pm 36$    &\nodata & 10 \\
        V3885 Sgr      & 296.91905           & $-42.00751$           & 128.6 $^{+0.6}_{-0.5}$ & 4.8$\pm0.5\times10^{-12}$*        & 6.1$\pm0.6\times10^{-12}$*    & 0.43 – 2.1       & 256 $\pm 26$*   & 9  & 10 \\ 
        V5662 Sgr      & 301.46284           & $-29.58358$           & 169$\pm$4              & 3.3$\pm$0.9$\times10^{-12}$       & 4.2$\pm$1.1$\times10^{-12}$   & 0.43 – 2.1       & < 24            & 3  & 10 \\
        \textbf{SS Cyg}& 325.67904           & \phantom{$-$}43.58622 & 112.3 $^{+0.4}_{-0.3}$ & 9.0$\pm0.9\times10^{-10}$*        & \nodata                       & 4.6              & 180 $\pm 20$*   & 16 & 16 \\ 
        \nodata        & \nodata             & \nodata               & \nodata                & 4.1$^{+0.5}_{-0.4}\times10^{-11}$*&  \nodata                      & \nodata          & 810 $\pm 81$*   &\nodata&\nodata\\ 
        \hline
    \end{tabular}
\end{table*}


\section*{Acknowledgements}

COH is supported by NSERC Discovery Grant RGPIN-2016-04602.
GRS and AJH are supported by NSERC Discovery Grant RGPIN-2021-04001. 
The National Radio Astronomy Observatory is a facility of the National Science Foundation operated under cooperative agreement by Associated Universities, Inc.
This research has made use of the CIRADA cutout service at URL cutouts.cirada.ca, operated by the Canadian Initiative for Radio Astronomy Data Analysis (CIRADA). CIRADA is funded by a grant from the Canada Foundation for Innovation 2017 Innovation Fund (Project 35999), as well as by the Provinces of Ontario, British Columbia, Alberta, Manitoba and Quebec, in collaboration with the National Research Council of Canada, the US National Radio Astronomy Observatory and Australia’s Commonwealth Scientific and Industrial Research Organisation.
This research also made use of TOPCAT \citep{topcat}, software designed to handle large astronomical datasets with virtual observatory tools.

\section*{Data Availability}

VLASS Quicklook data, including those used for this paper, are available via \url{https://cirada.ca/vlasscatalogueql0}.


\bibliographystyle{mnras}
\bibliography{references}


\bsp	
\label{lastpage}
\end{document}